\documentclass[a4paper,12pt]{article}
\usepackage{epsfig}
\usepackage[dvips,usenames]{color}
\usepackage{graphicx}

\newlength{\dinwidth}
\newlength{\dinmargin}
\begin{document}
\title{ Polarization Anomaly  in $B\to \phi K^{*}$ \\
and Probe of Tensor Interactions }
\bigskip

\author{C.~S. Kim\footnote{cskim@yonsei.ac.kr}
\\
{ \small \it Department of Physics and IPAP, Yonsei University,
Seoul 120-479, Korea} \\
{\small \it Physics Division, National Center for Theoretical Sciences, Hsinchu 300, Taiwan
}\\
\\
Y.~D. Yang\footnote{yangyd@henannu.edu.cn}
\\
{ \small \it Department of Physics, Henan Normal University,
Xinxiang, Henan 453002, P.R. China
}\\
} \maketitle
\begin{picture}(0,0)
\end{picture}

\begin{abstract}

\noindent The pure penguin process $B\to \phi K^*$ is one of the
most important probes of physics beyond the Standard Model.
Recently BaBar and Belle have measured the unexpectedly large
transverse polarization in the decays $B\to \phi K^*$, which may
single out new physics effects beyond the Standard model. We study
the possibility that the phenomenon could serve as an important
probe of anomalous tensor interactions. We find that a spin flipped
tensor interaction with a small strength and a phase could give a
possible solution to the polarization puzzle.
\\
{\bf PACS Numbers 13.25.Hw,  12.60.Jv}
\end{abstract}

\newpage
Looking for signals of physics beyond the Standard Model is one of
the most important missions of high energy physics. It is well
known that flavor-changing neutral currents induced in $B$ decays are one of the best
probes of new physics beyond the Standard Model because they
arise only through loop effects in the Standard Model (SM). To this end, the
decays $B\to \phi K^*$ are of particular
interests, since they are pure penguin processes and have
interesting polarization  phenomena as well as relatively clear experiment
signature. Within the SM, it is expected that both $\phi$ and $K^*$
are mainly longitudinally polarized, while its transverse
polarization is suppressed by the power of $m_{\phi,K^*}/m_{B}$.
However, last year both BaBar and Belle had observed rather small
longitudinal polarizations in the decays
\begin{eqnarray}
f_{L}(\phi K^{*+})&=&0.46\pm 0.12\pm 0.03~,~~~ f_{L}(\phi
K^{*0})=0.65\pm 0.07\pm 0.02~~~~  ({\rm BaBar}~~\cite{babar03}),
\\
f_{L}(\phi K^{*0})&=&0.43\pm 0.09\pm 0.04~,~~~ f_{T}(\phi
K^{*0})=0.41\pm 0.10\pm 0.04~~~~ ({\rm Belle}~~\cite{belle03}).
\end{eqnarray}
Due to $ |f_{0}|+|f_{T}|=1$, both groups have measured unexpectedly
large transverse polarizations in the $B\to \phi K^{\ast}$ decays.

This summer BaBar Collaboration has again reported their full angular
analysis of the the decay $B^{0}\to\phi K^{*}$ ~\cite{babar04}
\begin{equation}
f_{L}(\phi K^{*0})=0.52\pm 0.05\pm 0.02~, ~~~ f_{T}(\phi
K^{*0})=0.22\pm 0.05\pm 0.02~,
\end{equation}
which has confirmed their previous measurements and called urgent theoretical
explanations.

The final states $\phi$ and $K^{\ast}$ are fast moving in the $B$
meson frame and any spin flip of fast flying quark will be
suppressed by power of $m_{q}/E$. The charge interaction currents
structure of the SM is left-handed, therefore, will result in  the
dominance of longitudinal polarization. Such situation has been
known to us for many years~\cite{ali,korner, grossman}. So that,
the recent measurements of large transverse polarizations in
$B\to \phi K^*$ are referred as a puzzle
within the high energy physics community~\cite{talks}.

The analysis of the decays within the SM can be performed in terms of an
effective low-energy theory  with the Hamiltonian
\cite{buras}
\begin{equation}
\label{heff} {\cal H}_{\rm eff}^{\rm SM} =-\frac{G_{F}}{\sqrt{2}} V_{tb}
V_{ts}^* \sum_{i=3}^{10} C_{i}O_{i}~.
\end{equation}
The amplitude for the decay within the SM can be written as
\begin{eqnarray}
\mathcal{A}_{\lambda\lambda}=&& -\frac{G_F}{\sqrt{2}}V_{tb}V_{ts}^*
a(\phi K^*) \langle \phi |\bar{s}\gamma_{\mu}s|0\rangle \langle
K^*
| \bar{s}\gamma^{\mu} (1-\gamma_{5})b| B\rangle \nonumber \\
 = && \frac{G_F}{\sqrt{2}}V_{tb}V_{ts}^{*}a(\phi K^* )i
 f_{\phi}m_{\phi}\left[ \varepsilon^{*}_{1}\cdot \varepsilon^{*}_{2}
 (M_{B}+M_{K^{*}})A_{1}(m_{\phi}^{2})
 \right.
 \nonumber \\
&&
 \left.
 -(\varepsilon^{*}_{1}\cdot P_{B})(\varepsilon^{*}_{2}\cdot P_{B})
  \frac{2A_{2}(m_{\phi}^2 ) }{M_B +M_{K^*}}
  +i \epsilon_{\mu\nu\alpha\beta} \varepsilon^{*\mu}_{2}
  \varepsilon^{*\nu}_{1}P_{B}^{\alpha}P_{K^*}^{\sigma}
  \frac{2 V(m_{\phi}^2)}{M_B +M_{K^* }}
   \right]~.
\end{eqnarray}
Since $B$ meson is a pseudoscalar, the final two vector mesons
must have the same helicity. In the helicity basis, the amplitude
can be decomposed into three helicity amplitudes, which are
\begin{eqnarray}
H_{00}&=&\frac{G_F}{\sqrt{2}}V_{tb}V_{ts}^{*}a(\phi K^*
)\frac{if_{\phi}}{2M_{K^* }}
 \left[ (M_{B}^2 -M_{K^*}^2 -M_{\phi}^2 ) (M_B +M_{K^* })A_{1}
  -\frac{4M_{B}^2 p_c^2 A_{2}}{M_{B}+M_{K^* }}
 \right]~,\nonumber \\
 H_{\pm\pm}&=&i\frac{G_F}{\sqrt{2}}V_{tb}V_{ts}^{*}
 a(\phi K^* )M_{\phi}f_{\phi} \left[
 (M_{B}+M_{K^*})A_1 \mp \frac{2M_B p_c }{M_B +M_{K^*}}V
 \right]~.
\end{eqnarray}
In naive factorization~~\cite{cheng},
$$a(\phi K^*)=a_3 +a_4 +a_5 -\frac{1}{2}(a_7 +a_9 +a_{10} )~.$$
Then the branching ratio is thus read as
\begin{equation}
\mathcal{B}=\frac{\tau_B p_c }{8\pi M^{2}_B}
 \left( |H_{0}|^2 +|H_{+}|^2 |H_{-}|^2
 \right)~.
\end{equation}
And the longitudinal and the transverse polarization rates are
\begin{equation}
f_{L}=\frac{\Gamma_L}{\Gamma}=\frac{|H_{0}|^2 }{|H_{0}|^2
+|H_{+}|^2 |H_{-}|^2}~, ~~
f_{T}=\frac{\Gamma_T}{\Gamma}=\frac{|H_{+}|^2 +|H_{-}|^2
}{|H_{0}|^2 +|H_{+}|^2 +|H_{-}|^2}~.
\end{equation}
Using the Wilson coefficients $c_{3-6}$ evaluated at scale of
$\mu =m_b$~\cite{buras}, and the decay constants $f_B = 0.18$ GeV,
$f_{\phi} = 0.221$ GeV and the form factors of light-cone QCD
sum-rules~\cite{pball}, one can get
$$
\mathcal{B}\mathbf{}(B\to \phi K^* )\sim 8.55\times  10^{-6}~, ~~~
f_{L}\sim 0.90~,~~~ f_{T}\sim 0.09~.
$$

It must be reminded that a theoretical estimation of the branching ratios
depend very strongly on the form factors from different hadronic models and the
theoretical frameworks of $B$ meson nonletponic decays, even though most
frameworks and form factors predict dominance of the longitudinal
polarizations. For example, recent calculation of
$Br(B^0 \to \phi K^{\ast0})$ by Cheng and Yang by using QCD
factorization~\cite{BBNS} gives $Br \sim 8.71\times 10^{-6}$ for LCSR and $4.62
\times 10^{-6}$ for BSW form-factors~\cite{pball, BSW},
respectively, while pQCD~\cite{lihn} calculation gives $Br \sim 14.86 \times
10^{-6}$ where form-factors are not inputs. However, both studies
present the dominance of longitudinal polarization.

After BaBar and Belle measurements of the abnormal large
transverse  polarization, there have been some theoretical explanations;
namely, through the final state interactions (FSI) contributions~\cite{FSI},
large annihilation contributions and new physics from right-hand
current interactions~\cite{kagan} and transverse $\phi$ from the
emitted gluon of $b\to s g^{\ast}$ which might be enhanced by new
physics~\cite{houws}.

In this letter, we investigate the possibility that the abnormally
large transverse polarization may arise from a new tensor
interaction beyond the SM (bSM),
\begin{equation}\label{tensor}
  \mathcal{H}_{\rm eff}^{\rm bSM}=\frac{G_F}{\sqrt{2}}|V_{ts}|g_{_T}e^{i\delta_T}
  \overline{s}\sigma_{\mu\nu}(1+\gamma_5 )s\otimes\bar s
  \sigma^{\mu\nu}(1+\gamma^5 )b~,
\end{equation}
where $g_{_T}$ is the relative interaction strength normalized to
that of $b\to s\bar{s}s$ in the SM and $i\delta_T$ is the new
physics phase. In principle, such a tensor operator could be
produced even in MSSM~\cite{csh,hiller}. Interestingly, the recent
study of radiative pion decay $\pi^+ \to e^+ \nu \gamma$ at
PIBETA~\cite{pibeta} has found deviations from the SM in the
high-$E_{\gamma}$ kinematic region, which may indicate the
existence of a tensor quark-lepton interaction~\cite{pob,chizhov}.
We also note that Kagan  mentioned the case of tensor operator for
resolving the puzzle~\cite{kagan}.

Our starting point arises from the observation that the tensor
interaction only contributes to transverse polarization but not to
longitudinal one. The matrix element reads~\cite{pball}
\begin{equation}\label{ft}
\langle \phi(q,\epsilon^{T*} )| \bar{s}\sigma_{\mu\nu}s|0\rangle=
-i f_{\phi}^{T}(\epsilon^{T*}_{\mu}q_{\nu} -
q_{\mu}\epsilon^{T*}_{\nu})~,
\end{equation}
which is scaled  as $f^{T} E_{\phi} $ since $q\sim E_{\phi}(1,0,0,
1)$ for fast flying $\phi$. However, if the $\phi$ meson is
produced instead from a vector interaction vertex, we will have
$\langle \phi(q,\epsilon)| \bar{s}\gamma_{\mu}s|0\rangle =f_{\phi}
m_{\phi}\epsilon^{*}_{\mu}$, and it is easy to understand that the
longitudinal polarization dominate over the transverse one by a
large factor $m_{B}/m_{\phi}$ because of $\epsilon_{L}^{\mu}\to
q^{\mu}/m_{\phi}$.

Using the form factors defined in Ref.~\cite{pball}, we can write
down the amplitude of the tensor operator in Eq. (\ref{tensor}) in
naive factorization approximation,
\begin{eqnarray}
&& \langle \phi(q) K^{*}(p)|\mathcal{H}_{eff}^{T} |B(p_{B})\rangle
=\frac{G_F}{\sqrt{2}}|V_{ts}|g_{T}e^{i\delta_T}(-2if_{\phi}^{T}) \times
\nonumber \\
&& ~~~~~ \{ \epsilon^{T*}_{\phi}\cdot \epsilon^{T*}_{K^*}
T_{2}(m^2_{\phi}) (m^2_B - m^2_{K^*})+2i
T_{1}(m^2_{\phi})\varepsilon_{\mu\nu\alpha\beta}\epsilon^{T*\mu}_{\phi}
\epsilon^{T*\nu}_{K^*} p^{\alpha}_{B}p^{\beta} \}~.
\end{eqnarray}
{}From this equation, we can get the new physics contributions,
\begin{eqnarray}
H_{00}^{\rm bSM}&=&0~,\\
H_{\pm\pm}^{\rm bSM}&=&\frac{G_F}{\sqrt{2}}|V_{ts}|g_{_T}e^{i\delta_T}(-2if_{\phi}^{T})
\left[ (m^2_B -m^2_{K^*})T_{2}(m^2_{\phi})\mp 2 m_{B}p_c
T_{1}(m^2_{\phi})\right]~.
\end{eqnarray}
Compared with Eq.9, the tensor interaction contributions to
$H_{\pm\pm}$ are enhanced by a factor of $M_{B}/m_{\phi}$.
\begin{figure}
\begin{tabular}{cc}
\scalebox{0.7}{\epsfig{file=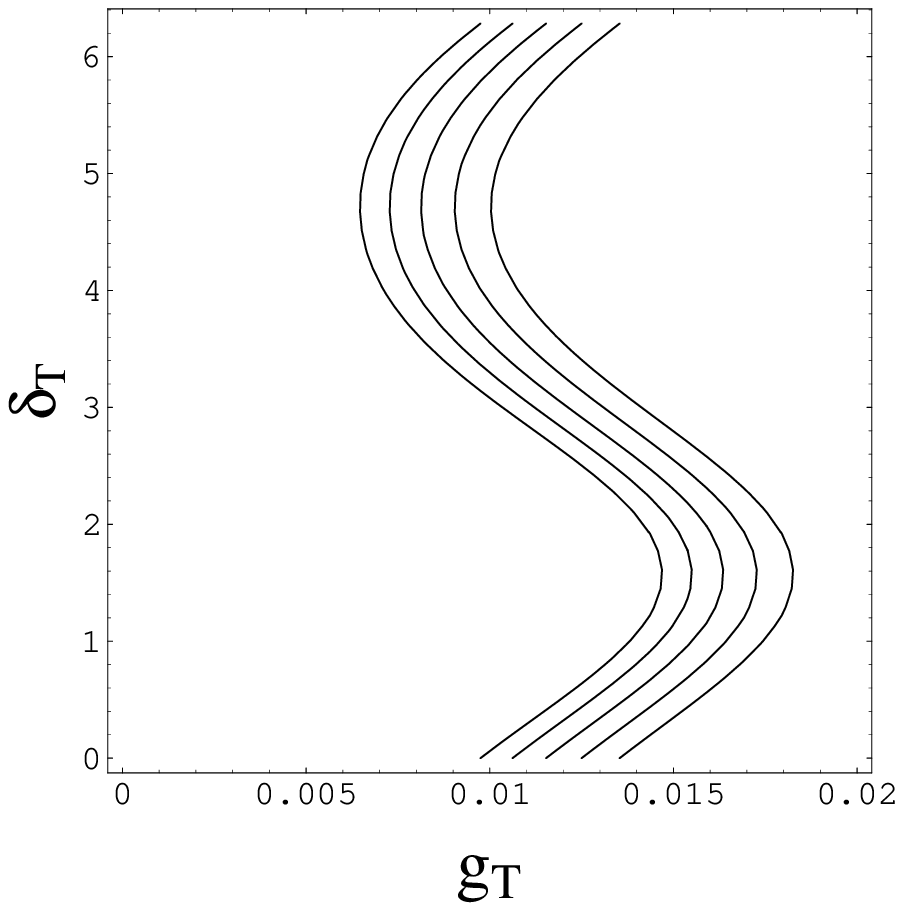}} & \scalebox{0.7}{\epsfig{file=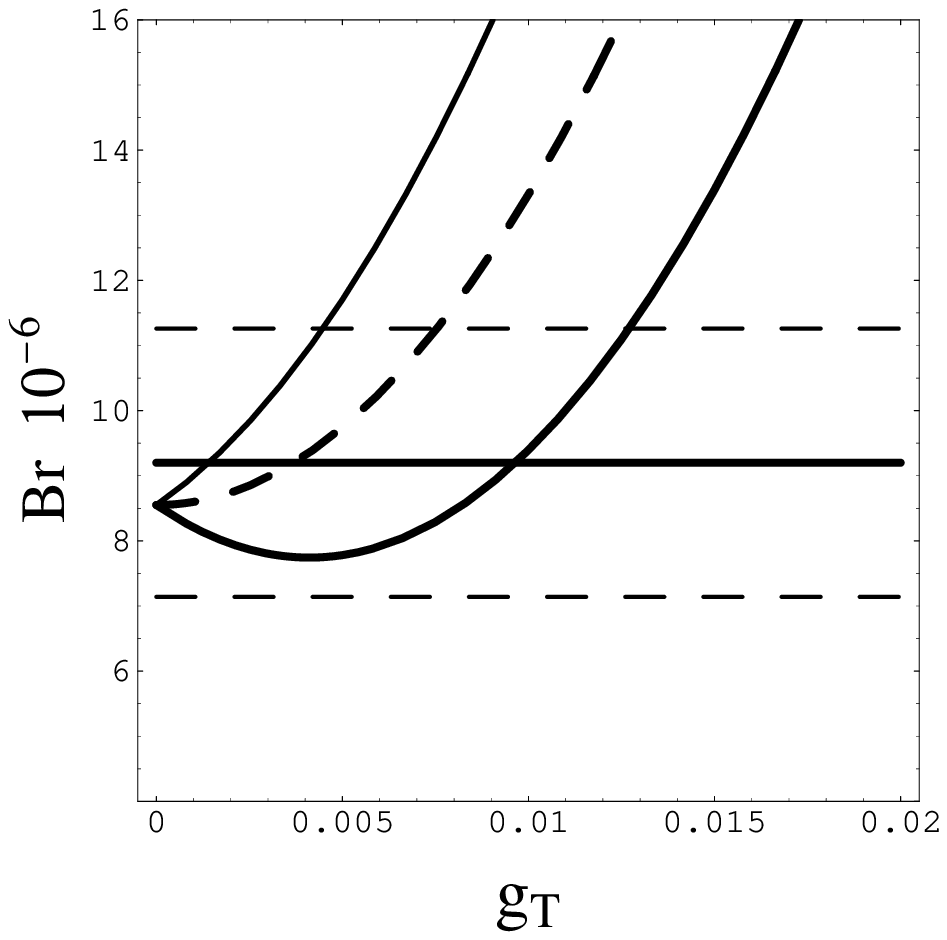}}  \\
(a)&(b)
\end{tabular}
\caption{\small (a) The contour plot for
$f_{L}=0.52\pm0.05\pm0.002$ within $2\sigma$. The central curve
is for the center value of $f_{L}$, and the
nearby curves are for $f_{L}$ at $1\sigma$, $2\sigma$ variances,
respectively. (b) The branching ratio constraints: The thin
curve, the dashed curve and the thick curve are for
$\delta_{T}=3\pi/2,~\pi,~\pi/2$, respectively. The horizontal lines
are the experimental results for $Br(B^0 \to \phi K^{*})$ with
$2\sigma$ variance. }
\end{figure}

Numerical results are presented in Fig. 1. From Fig. 1(a), we can
 find that the transverse polarization in $B^{0}\to \phi K^{*0}$
 is very sensitive to the presence of new tensor interactions. For
 $ 0.005<g_{_T}<0.02 $, we can easily find solutions to the polarization
 puzzle depending on the phase of the tensor interaction. For
 example, to account for $f_{L}=0.52\pm0.05\pm0.002$ within $2\sigma$,
 we get intervals $g_{_T}\in (0.014,0.019)$, $(0.009,0.015)$, $(0.006,0.011)$
 for $\delta_{T}=\pi/2, \pi, 3\pi/2$, respectively.
 Of course, the branching ratio measurements
 could also give constraints on such a tensor interaction operator,
 which are presented in Fig. 1(b). Here we can see the windows are
 very narrow because the longitudinal contribution estimated within the
 SM already saturate the experimental branching ratio. However,
 it is well known that theoretical calculations of the branching ratios
 of hadronic $B$ decays suffer from large uncertainties. It is
 believed that polarization fractions could be predicted more
 accurately than the branching ratios, because some of hadronic
 uncertainties could be cancelled in the former ones.  In the
 future, if theoretical frameworks for hadronic $B$ decays could
 achieve $10\%$ accuracy and their predictions of longitudinal
 branching ratio still saturate the experimental measurement,
 the tensor interaction scenario could be ruled out.
 In such a case, we need not only new physics contributions to
 transverse part but also new contributions destructive to
 longitudinal part. However, it would be very hard to account for
 the large branching ratio of $B\to \phi K$ because the similarity
 between the amplitudes of $B\to \phi K$ and the longitudinal
 amplitude of  $B\to \phi K^{*}$ in the heavy $M_b$ limit.

In conclusion, we have studied the large transverse polarization
puzzle in $B\to \phi K^{*}$ decays, which is taken as an important
probe of an anomalous tensor interactions. We find that a
relatively weak tensor interaction could resolve the puzzle. If we
take the coupling $g_{_T}=m^2_{W}/\Lambda^2_{T}$, such a solution
might be a signal of new physics with tensor interaction at TeV
scale.  With the running of $B$ factories BaBar and Belle, we have
witnessed many challenging phenomena. Theoretically, we need more
accurate and complete framework to clarify whether the SM could
explain those abnormal phenomena or not.
\\

{\it Note added: When we finished our work, we note the
paper\cite{kcyang} where the same tensor operator is also studied.
}

\section*{Acknowledgments}

The work of C.S.K. was supported
by Grant No. R02-2003-000-10050-0 from BRP of the KOSEF.
Y.D. is supported in part by NFSC under contract No.10305003, Henan
Provincial Foundation for Prominent Young Scientists under
contract No.0312001700 and in part by the project sponsored by SRF for
ROCS, SEM. This work is also supported by Grant No.
F01-2004-000-10292-0 of KOSEF-NSFC International
Collaborative Research Grant.

\end{document}